\documentclass[aps,prd,twocolumn,superscriptaddress]{revtex4-2}
\usepackage{amssymb}
\usepackage{color}
\usepackage{multirow}
\usepackage{lineno}
\usepackage[dvipdfmx]{graphicx}
\newcommand{\ibd}{\overline{\nu}_{\text{e}}\,+\,\text{p}\,\to\,\text{e}^{+}\,+\,\text{n}}

\begin{document}

% Use the \preprint command to place your local institutional report
% number in the upper righthand corner of the title page in preprint mode.
% Multiple \preprint commands are allowed.
% Use the 'preprintnumbers' class option to override journal defaults
% to display numbers if necessary
%\preprint{}

%Title of paper
\title{Measurement of the neutrino-oxygen neutral-current quasielastic cross section\\using atmospheric neutrinos in the SK-Gd experiment}

% repeat the \author .. \affiliation  etc. as needed
% \email, \thanks, \homepage, \altaffiliation all apply to the current
% author. Explanatory text should go in the []'s, actual e-mail
% address or url should go in the {}'s for \email and \homepage.
% Please use the appropriate macro foreach each type of information

% \affiliation command applies to all authors since the last
% \affiliation command. The \affiliation command should follow the
% other information
% \affiliation can be followed by \email, \homepage, \thanks as well.
%\author{}
%\email[]{Your e-mail address}
%\homepage[]{Your web page}
%\thanks{}
%\altaffiliation{}
%\affiliation{}

%Collaboration name if desired (requires use of superscriptaddress
%option in \documentclass). \noaffiliation is required (may also be
%used with the \author command).
%\collaboration can be followed by \email, \homepage, \thanks as well.
%\collaboration{}
%\noaffiliation
%\include{Authors/authors-20230602-orcid}
%\include{Authors/authors-20230602}
\newcommand{\AFFicrr}{\affiliation{Kamioka Observatory, Institute for Cosmic Ray Research, University of Tokyo, Kamioka, Gifu 506-1205, Japan}}
\newcommand{\AFFkashiwa}{\affiliation{Research Center for Cosmic Neutrinos, Institute for Cosmic Ray Research, University of Tokyo, Kashiwa, Chiba 277-8582, Japan}}
\newcommand{\AFFipmu}{\affiliation{Kavli Institute for the Physics and
Mathematics of the Universe (WPI), The University of Tokyo Institutes for Advanced Study,
University of Tokyo, Kashiwa, Chiba 277-8583, Japan }}
\newcommand{\AFFmad}{\affiliation{Department of Theoretical Physics, University Autonoma Madrid, 28049 Madrid, Spain}}
\newcommand{\AFFubc}{\affiliation{Department of Physics and Astronomy, University of British Columbia, Vancouver, BC, V6T1Z4, Canada}}
\newcommand{\AFFbu}{\affiliation{Department of Physics, Boston University, Boston, MA 02215, USA}}
\newcommand{\AFFuci}{\affiliation{Department of Physics and Astronomy, University of California, Irvine, Irvine, CA 92697-4575, USA }}
\newcommand{\AFFcsu}{\affiliation{Department of Physics, California State University, Dominguez Hills, Carson, CA 90747, USA}}
\newcommand{\AFFcnm}{\affiliation{Institute for Universe and Elementary Particles, Chonnam National University, Gwangju 61186, Korea}}
\newcommand{\AFFduke}{\affiliation{Department of Physics, Duke University, Durham NC 27708, USA}}
\newcommand{\AFFgifu}{\affiliation{Department of Physics, Gifu University, Gifu, Gifu 501-1193, Japan}}
\newcommand{\AFFgist}{\affiliation{GIST College, Gwangju Institute of Science and Technology, Gwangju 500-712, Korea}}
\newcommand{\AFFuh}{\affiliation{Department of Physics and Astronomy, University of Hawaii, Honolulu, HI 96822, USA}}
\newcommand{\AFFicl}{\affiliation{Department of Physics, Imperial College London , London, SW7 2AZ, United Kingdom }}
\newcommand{\AFFkek}{\affiliation{High Energy Accelerator Research Organization (KEK), Tsukuba, Ibaraki 305-0801, Japan }}
\newcommand{\AFFkobe}{\affiliation{Department of Physics, Kobe University, Kobe, Hyogo 657-8501, Japan}}
\newcommand{\AFFkyoto}{\affiliation{Department of Physics, Kyoto University, Kyoto, Kyoto 606-8502, Japan}}
\newcommand{\AFFliv}{\affiliation{Department of Physics, University of Liverpool, Liverpool, L69 7ZE, United Kingdom}}
\newcommand{\AFFmiyagi}{\affiliation{Department of Physics, Miyagi University of Education, Sendai, Miyagi 980-0845, Japan}}
\newcommand{\AFFnagoya}{\affiliation{Institute for Space-Earth Environmental Research, Nagoya University, Nagoya, Aichi 464-8602, Japan}}
\newcommand{\AFFkmi}{\affiliation{Kobayashi-Maskawa Institute for the Origin of Particles and the Universe, Nagoya University, Nagoya, Aichi 464-8602, Japan}}
\newcommand{\AFFpol}{\affiliation{National Centre For Nuclear Research, 02-093 Warsaw, Poland}}
\newcommand{\AFFsuny}{\affiliation{Department of Physics and Astronomy, State University of New York at Stony Brook, NY 11794-3800, USA}}
\newcommand{\AFFokayama}{\affiliation{Department of Physics, Okayama University, Okayama, Okayama 700-8530, Japan }}
\newcommand{\AFFosaka}{\affiliation{Department of Physics, Osaka University, Toyonaka, Osaka 560-0043, Japan}}
\newcommand{\AFFox}{\affiliation{Department of Physics, Oxford University, Oxford, OX1 3PU, United Kingdom}}
\newcommand{\AFFqmul}{\affiliation{School of Physics and Astronomy, Queen Mary University of London, London, E1 4NS, United Kingdom}}
\newcommand{\AFFregina}{\affiliation{Department of Physics, University of Regina, 3737 Wascana Parkway, Regina, SK, S4SOA2, Canada}}
\newcommand{\AFFseoul}{\affiliation{Department of Physics, Seoul National University, Seoul 151-742, Korea}}
\newcommand{\AFFsheff}{\affiliation{Department of Physics and Astronomy, University of Sheffield, S3 7RH, Sheffield, United Kingdom}}
\newcommand{\AFFshizuokasc}{\affiliation{Department of Informatics in
Social Welfare, Shizuoka University of Welfare, Yaizu, Shizuoka, 425-8611, Japan}}
\newcommand{\AFFstfc}{\affiliation{STFC, Rutherford Appleton Laboratory, Harwell Oxford, and Daresbury Laboratory, Warrington, OX11 0QX, United Kingdom}}
\newcommand{\AFFskk}{\affiliation{Department of Physics, Sungkyunkwan University, Suwon 440-746, Korea}}
\newcommand{\AFFtodai}{\affiliation{Department of Physics, University of Tokyo, Bunkyo, Tokyo 113-0033, Japan }}
\newcommand{\AFFtit}{\affiliation{Department of Physics,Tokyo Institute of Technology, Meguro, Tokyo 152-8551, Japan }}
\newcommand{\AFFtus}{\affiliation{Department of Physics, Faculty of Science and Technology, Tokyo University of Science, Noda, Chiba 278-8510, Japan }}
\newcommand{\AFFtoronto}{\affiliation{Department of Physics, University of Toronto, ON, M5S 1A7, Canada }}
\newcommand{\AFFtriumf}{\affiliation{TRIUMF, 4004 Wesbrook Mall, Vancouver, BC, V6T2A3, Canada }}
\newcommand{\AFFtokai}{\affiliation{Department of Physics, Tokai University, Hiratsuka, Kanagawa 259-1292, Japan}}
\newcommand{\AFFtsinghua}{\affiliation{Department of Engineering Physics, Tsinghua University, Beijing, 100084, China}}
\newcommand{\AFFynu}{\affiliation{Department of Physics, Yokohama National University, Yokohama, Kanagawa, 240-8501, Japan}}
\newcommand{\AFFllr}{\affiliation{Ecole Polytechnique, IN2P3-CNRS, Laboratoire Leprince-Ringuet, F-91120 Palaiseau, France }}
\newcommand{\AFFbari}{\affiliation{ Dipartimento Interuniversitario di Fisica, INFN Sezione di Bari and Universit\`a e Politecnico di Bari, I-70125, Bari, Italy}}
\newcommand{\AFFnapoli}{\affiliation{Dipartimento di Fisica, INFN Sezione di Napoli and Universit\`a di Napoli, I-80126, Napoli, Italy}}
\newcommand{\AFFroma}{\affiliation{INFN Sezione di Roma and Universit\`a di Roma ``La Sapienza'', I-00185, Roma, Italy}}
\newcommand{\AFFpadova}{\affiliation{Dipartimento di Fisica, INFN Sezione di Padova and Universit\`a di Padova, I-35131, Padova, Italy}}
\newcommand{\AFFkeio}{\affiliation{Department of Physics, Keio University, Yokohama, Kanagawa, 223-8522, Japan}}
\newcommand{\AFFwinnipeg}{\affiliation{Department of Physics, University of Winnipeg, MB R3J 3L8, Canada }}
\newcommand{\AFFkcl}{\affiliation{Department of Physics, King's College London, London, WC2R 2LS, UK }}
\newcommand{\AFFwarwick}{\affiliation{Department of Physics, University of Warwick, Coventry, CV4 7AL, UK }}
\newcommand{\AFFral}{\affiliation{Rutherford Appleton Laboratory, Harwell, Oxford, OX11 0QX, UK }}
\newcommand{\AFFwu}{\affiliation{Faculty of Physics, University of Warsaw, Warsaw, 02-093, Poland }}
\newcommand{\AFFbcit}{\affiliation{Department of Physics, British Columbia Institute of Technology, Burnaby, BC, V5G 3H2, Canada }}
\newcommand{\AFFtohoku}{\affiliation{Department of Physics, Faculty of Science, Tohoku University, Sendai, Miyagi, 980-8578, Japan }}
\newcommand{\AFFicise}{\affiliation{Institute For Interdisciplinary Research in Science and Education, ICISE, Quy Nhon, 55121, Vietnam }}
\newcommand{\AFFilance}{\affiliation{ILANCE, CNRS - University of Tokyo International Research Laboratory, Kashiwa, Chiba 277-8582, Japan}}
\newcommand{\AFFibs}{\affiliation{Institute for Basic Science (IBS), Daejeon, 34126, Korea}}
\newcommand{\AFFglasgow}{\affiliation{School of Physics and Astronomy, University of Glasgow, Glasgow, Scotland, G12 8QQ, United Kingdom}}
\newcommand{\AFFoecu}{\affiliation{Media Communication Center, Osaka Electro-Communication University, Neyagawa, Osaka, 572-8530, Japan}}

\AFFicrr
\AFFkashiwa
\AFFmad
\AFFbu
\AFFbcit
\AFFuci
\AFFcsu
\AFFcnm
\AFFduke
\AFFllr
\AFFgifu
\AFFgist
\AFFglasgow
\AFFuh
\AFFibs
\AFFicise
\AFFicl
\AFFbari
\AFFnapoli
\AFFpadova
\AFFroma
\AFFilance
\AFFkeio
\AFFkek
\AFFkcl
\AFFkobe
\AFFkyoto
\AFFliv
\AFFmiyagi
\AFFnagoya
\AFFkmi
\AFFpol
\AFFsuny
\AFFokayama
\AFFoecu
\AFFox
\AFFral
\AFFseoul
\AFFsheff
\AFFshizuokasc
\AFFstfc
\AFFskk
\AFFtohoku
\AFFtokai
%\AFFtokyo
\AFFtodai
\AFFipmu
\AFFtit
\AFFtus
%\AFFtoronto
\AFFtriumf
\AFFtsinghua
\AFFwu
\AFFwarwick
\AFFwinnipeg
\AFFynu

\author{S.~Sakai}
\AFFokayama

%%%%%%%%%%%%%%%%%%%%%%%%%%%%%%%%%%%%%%%%%%%%%%%%%%%%%%%%%%%%%%%%%%%%
%ICRR
\author{K.~Abe}
\AFFicrr
\AFFipmu
\author{C.~Bronner}
\AFFicrr
\author{Y.~Hayato}
\AFFicrr
\AFFipmu
\author{K.~Hiraide}
\AFFicrr
\AFFipmu
\author{K.~Hosokawa}
\AFFicrr
\author{K.~Ieki}
\author{M.~Ikeda}
\AFFicrr
\AFFipmu
\author{J.~Kameda}
\AFFicrr
\AFFipmu
\author{Y.~Kanemura}
\author{R.~Kaneshima}
\author{Y.~Kashiwagi}
\AFFicrr
% ...

%%%%%%%%%%%%%%%%%%%%%%%%%%%%%%%%%%%%%%%%%%%%%%%%%%%%%%%%%%%%%%%%%%%%%

\collaboration{The Super-Kamiokande Collaboration}
\noaffiliation

\date{\today}

\begin{abstract}
% insert abstract here
We report the first measurement of the atmospheric neutrino-oxygen neutral-current quasielastic (NCQE) cross section in the gadolinium-loaded Super-Kamiokande (SK) water Cherenkov detector.
In June 2020, SK began a new experimental phase, named SK-Gd, by loading 0.011\% by mass of gadolinium into the ultrapure water of the SK detector.
The introduction of gadolinium to ultrapure water has the effect of improving the neutron-tagging efficiency.
Using a 552.2 day data set from August 2020 to June 2022, we measure the NCQE cross section to be 0.74\,$\pm$\,0.22(stat.)\,$^{+\text{0.85}}_{-\text{0.15}}$(syst.)\,$\times$\,10$^{-\text{38}}$\,cm$^{\text{2}}$$/$oxygen in the energy range from 160~MeV to 10~GeV, which is consistent with the atmospheric neutrino-flux-averaged theoretical NCQE cross section and the measurement in the SK pure-water phase within the uncertainties.
Furthermore, we compare the models of the nucleon-nucleus interactions in water and find that the Binary Cascade model and the Li$\grave{\text{e}}$ge Intranuclear Cascade model provide a somewhat better fit to the observed data than the Bertini Cascade model.
Since the atmospheric neutrino-oxygen NCQE reactions are one of the main backgrounds in the search for diffuse supernova neutrino background (DSNB), these new results will contribute to future studies -- and the potential discovery -- of the DSNB in SK.
\end{abstract}

% insert suggested keywords - APS authors don't need to do this
%\keywords{}

%\maketitle must follow title, authors, abstract, and keywords
\maketitle

% body of paper here - Use proper section commands
% References should be done using the \cite, \ref, and \label commands
% Put \label in argument of \section for cross-referencing
%\section{\label{}}

\section{INTRODUCTION}\label{INTROD}
Neutrinos emitted from all past core-collapse supernovae comprise an integrated flux called the diffuse supernova neutrino background (DSNB)~\cite{2010Beacom}.
Detecting the DSNB would contribute to our understanding of the mechanism of supernova explosions and the history of star formation.
The Super-Kamiokande Gadolinium (SK-Gd) experiment~\cite{2004Beacom} is aiming to achieve the first observation of the DSNB.
In the DSNB search in a water Cherenkov experiment, we look for the inverse beta decay (IBD) events by electron antineutrinos ($\ibd$).
The positron emits Cherenkov photons immediately, while the neutron is thermalized and then captured on Gd, emitting a total of about 8~MeV of gamma-rays.
The gamma-rays give their energy to electrons or positrons via Compton scattering or pair production, then Cherenkov photons are emitted.
By detecting the positron signal (prompt signal) and the neutron signal (delayed signal), we can remove a large number of backgrounds that do not contain neutrons.
However, backgrounds that contain neutrons cannot be completely removed.\par
One of the main backgrounds in the DSNB search is caused by the atmospheric neutrino-oxygen neutral-current quasielastic (NCQE) reactions.
NCQE reactions can be expressed as
\begin{eqnarray}
%\nu(\bar{\nu})\,+\,^{\text{16}}\text{O}\,&\to&\,\nu(\bar{\nu})\,+\,^{\text{15}}\text{O}^{*}\,+\,\text{n}, \nonumber
\nu(\bar{\nu})\,+\,^{\text{16}}\text{O}\,&\to&\,\nu(\bar{\nu})\,+\,^{\text{15}}\text{O}\,+\,\gamma\,+\,\text{n}, \nonumber
\\
%\nu(\bar{\nu})\,+\,^{\text{16}}\text{O}\,&\to&\,\nu(\bar{\nu})\,+\,^{\text{15}}\text{N}^{*}\,+\,\text{p},
\nu(\bar{\nu})\,+\,^{\text{16}}\text{O}\,&\to&\,\nu(\bar{\nu})\,+\,^{\text{15}}\text{N}\,+\,\gamma\,+\,\text{p},
\end{eqnarray}
where the atmospheric neutrino knocks out a nucleon of the oxygen nucleus, and the residual nucleus may emit one or more de-excitation gamma-rays with a few MeV.
When a neutron is knocked out, the combination of de-excitation gamma-rays and neutron mimics the IBD event, making it difficult to distinguish between NCQE and IBD events.
Therefore, the precise estimation of NCQE events is essential for the DSNB discovery in SK-Gd.\par
To estimate the NCQE events precisely, the behavior of neutrons in water must be understood.
In IBD events, the outgoing neutron has at most a few MeV, while in NCQE events, the knocked-out neutron may have hundreds of MeV.
In the latter case, it can knock out other nucleons of oxygen nuclei in water, and additional de-excitation gamma-rays and neutrons are generated.
Therefore, it is crucial to understand the nucleon-nucleus interactions in water (secondary interactions).\par
The T2K experiment measured the accelerator neutrino-oxygen NCQE cross section with a large uncertainty mainly coming from the de-excitation gamma-rays by secondary interactions~\cite{2014Abe,2019Abe}.
In previous SK DSNB searches, the expected number of NCQE background events, which was scaled using the T2K NCQE cross section, had large systematic uncertainties (60--80\%)~\cite{2021Abe,2023Harada}.\par
In the past, the secondary interaction model based on the Bertini Cascade model (BERT)~\cite{2015Wright} was the only choice in SK.
However, now other secondary interaction models like the Binary Cascade model (BIC)~\cite{2004Folger} and the Li$\grave{\text{e}}$ge Intranuclear Cascade model (INCL++)~\cite{2013Boudard} can be employed and compared with data.
In this paper, we discuss the reproducibility of the observed data in each secondary interaction model using atmospheric neutrino events.
Then we report the first measurement of the atmospheric neutrino-oxygen NCQE cross section in the Gd-loaded SK water Cherenkov detector.
%So far, the GEANT3-based~\cite{1994Brun}} SK detector simulation has been used in physics analyses.
%Thus the secondary interaction model based on the Bertini Cascade model (BERT)~\cite{2015Wright} was the only choice.
%However, recently, the Geant4-based~\cite{2016Allison}} SK detector simulation~\cite{2019Harada}} was newly developed for the SK-Gd experiment.
%As a result, we can compare the observed data with the latest secondary interaction models.
% The outline of this paper is as follows:
% First, we introduce the SK experiment in Sec. \ref{SK}.
% Next, we explain the simulation used in this measurement, event reconstruction and selection in Sec. \ref{SIMULA} and \ref{RECONS}.
% The results of this measurement is reported in Sec. \ref{RESULT}.
% In Sec. \ref{SECOND}, we introduce the comparison among nucleon-nucleus interaction models, which is another main topic in this paper.
% Lastly, we discuss the future prospects in Sec. \ref{FUTURE} before concluding in Sec. \ref{CONCLU}.

\section{THE SUPER-KAMIOKANDE EXPERIMENT}\label{SK}
The SK experiment~\cite{2003Fukuda} is located in Kamioka, Gifu, Japan, with the large water Cherenkov detector placed 1,000~m underground, resulting in a 2,700~m water equivalent overburden.
The rate of cosmic ray muons is reduced by a factor of 10$^{\text{5}}$ compared to that at ground level.
The detector consists of a stainless steel cylindrical water tank with a diameter of 39.3~m, a height of 42.0~m, containing 50~kilotons of ultrapure water.
The water tank is optically separated into an inner detector (ID) and an outer detector (OD).
The ID has 11,129 20-inch PMTs to reconstruct the energy, vertex position, direction, and kind of charged particles, while the OD has 1,885 8-inch PMTs to veto incoming cosmic ray muons.
Radioactive backgrounds are concentrated near the detector wall.
Thus events more than 2~m away from the ID wall are used in the analyses, resulting in a fiducial volume of 22.5~kilotons.\par
SK started its observation in April 1996, and so far, the observation is categorized into seven phases (from SK-I to SK-VII).
Since the start of SK-IV in 2008 we have been able to search for neutron signals up to 535~$\mu$s after the trigger thanks to an electronics upgrade~\cite{2009Watanabe,2009Nishino}.
However, through SK-V, the neutron signal was a 2.2~MeV gamma-ray from neutron capture on free protons (hydrogen in the water), and the neutron-tagging efficiency was low.
To increase the efficiency, we loaded 0.011\% by mass of Gd in SK, at which time the SK-VI (SK-Gd) phase started in July 2020~\cite{2022Abe}.
The time constant of neutron capture on Gd at this concentration is about 115~$\mu$s~\cite{2022Abe}.
Now 0.03\% of Gd has been loaded in SK, and we continue the observation as SK-VII since mid-2022.\par
The previous atmospheric neutrino-oxygen NCQE cross section measurement~\cite{2019Linyan} was performed using 2,778~days of SK-IV pure-water data from October 2008 to October 2017.
This study uses 552.2~days of SK-VI data with 0.011\% Gd-loaded water from August 2020 to June 2022.
This data set is the same as the one used for the DSNB search in SK-VI~\cite{2023Harada}.

\section{SIMULATION}\label{SIMULA}
The atmospheric neutrino flux at the SK detector is predicted using the HKKM11 model~\cite{2011Honda}, which shows good agreement with the observation in SK~\cite{2016Richard}.\par
Neutrino interactions are simulated using NEUT~\cite{2021Hayato} (version 5.4.0.1).
The NCQE cross section on oxygen is based on the model using the oxygen spectral function reported by Ankowski~\textit{et al}.~\cite{2005Benhar,2012Ankowski} with the BBBA05 vector form factor~\cite{2006Bradford} and the dipole axial form factor~\cite{2006Bradford}.
The state of the residual nucleus after neutrino-oxygen nucleus interaction (primary interaction) is selected based on the probabilities computed in Ref.~\cite{2012Ankowski}.
There are four states, $(p_{\text{1}/\text{2}})^{-\text{1}}$, $(p_{\text{3}/\text{2}})^{-\text{1}}$, $(s_{\text{1}/\text{2}})^{-\text{1}}$, and \textit{others}, where $($state$)^{-\text{1}}$ shows the state of the nucleus after a nucleon initially occupying the state ($p_{\text{1}/\text{2}}$, $p_{\text{3}/\text{2}}$, or $s_{\text{1}/\text{2}}$) is removed.
The production probability of each state is 0.1580, 0.3515, 0.1055, and 0.3850, respectively.
$(p_{\text{1}/\text{2}})^{-\text{1}}$ state is the ground state of $^{\text{15}}$O or $^{\text{15}}$N, thus no gamma-ray is emitted.
Mainly 6.18~MeV or 6.32~MeV gamma-rays are emitted from $(p_{\text{3}/\text{2}})^{-\text{1}}$ state of $^{\text{15}}$O or $^{\text{15}}$N, respectively~\cite{1994Leuschner,1991Ajzenberg}.
In the case of $(s_{\text{1}/\text{2}})^{-\text{1}}$ state, nucleons and gamma-rays are emitted because the excitation energy is high.
The de-excitation mode is selected based on the $^{\text{16}}\text{O}(\text{p}, \text{2p})$ experiment~\cite{2006Kobayashi}.
The \textit{others} state includes all other possibilities that are not in $(p_{\text{1}/\text{2}})^{-\text{1}}$, $(p_{\text{3}/\text{2}})^{-\text{1}}$, and $(s_{\text{1}/\text{2}})^{-\text{1}}$ states, and there are no data nor theoretical predictions covered by this state.
In our simulation, the \textit{others} state is set to be integrated into $(s_{\text{1}/\text{2}})^{-\text{1}}$ state by default.\par
In the past, a GEANT3-based~\cite{1994Brun} SK detector simulation where only BERT was implemented for neutron tracking in water was used.
However, a Geant4-based~\cite{2016Allison} (version 10.05.p01) SK detector simulation has been newly developed for the SK-Gd experiment.
In this simulation, BERT ({\tt \verb|FTFP_BERT_HP|} physics list), BIC ({\tt \verb|QGSP_BIC_HP|} physics list), and INCL++ ({\tt \verb|QGSP_INCLXX_HP|} physics list) can be used as the secondary interaction model.
Here, BERT is a traditional cascade model used in GEANT.
BIC uses a large set of hadron data to choose interaction processes to improve the accuracy.
INCL++ is an advanced binary cascade model including phase space and quantum mechanical processes.
The features of each secondary interaction model are described in Sec.~\ref{SECOND}.
In this NCQE cross section measurement, BERT is used as the baseline model.

\section{EVENT SELECTION}\label{RECONS}
In this study, we search for NCQE events that consist of prompt signals from de-excitation gamma-rays and delayed signals from neutrons.
We select the events where the visible energy of the prompt signal ($E_{\text{vis}}$) is between 7.49~MeV and 29.49~MeV because there are many NCQE events in this $E_{\text{vis}}$ region from simulation.
In each candidate event, delayed signals within 535 $\mu$s from the prompt signal are searched for.
The event reconstruction and neutron-tagging method follow the DSNB search in SK-Gd phase~\cite{2023Harada,2022Harada}.
The event reduction is described below.\par
First, several backgrounds in the energy range from a few MeV to tens of MeV are removed using the same reduction code as in the solar neutrino analysis~\cite{2016Abe}.
The removed background events are those caused by PMT noise hits, radioactive backgrounds, and decay electrons from cosmic ray muons.
These are removed by the goodness of reconstruction, the distance from the ID wall to the reconstructed vertex, and the time difference from preceding cosmic ray muons, respectively.
Second, the spallation events, which are prominent backgrounds in the energy range from a few MeV to tens of MeV, are removed.
These are decays of radioactive isotopes produced by nuclear spallation of oxygen nuclei induced by energetic cosmic ray muons, and are removed using the time and track of muons close to candidate events.
Moreover, atmospheric neutrino events other than NCQE events are removed.
For example, events associated with muons and pions are removed using two hit peaks close in time and the cleanliness of Cherenkov rings.
Finally, we select NCQE events using the reconstructed Cherenkov angle of the prompt signal ($\theta_{\text{C}}$) and the number of delayed signals per event ($N_{\text{delayed}}$).
Here, $\theta_{\text{C}}$ is determined from the distribution of angles calculated by a combination of 3-PMT hits.\par
\begin{figure}[tbp]
\includegraphics[width=8.6cm]{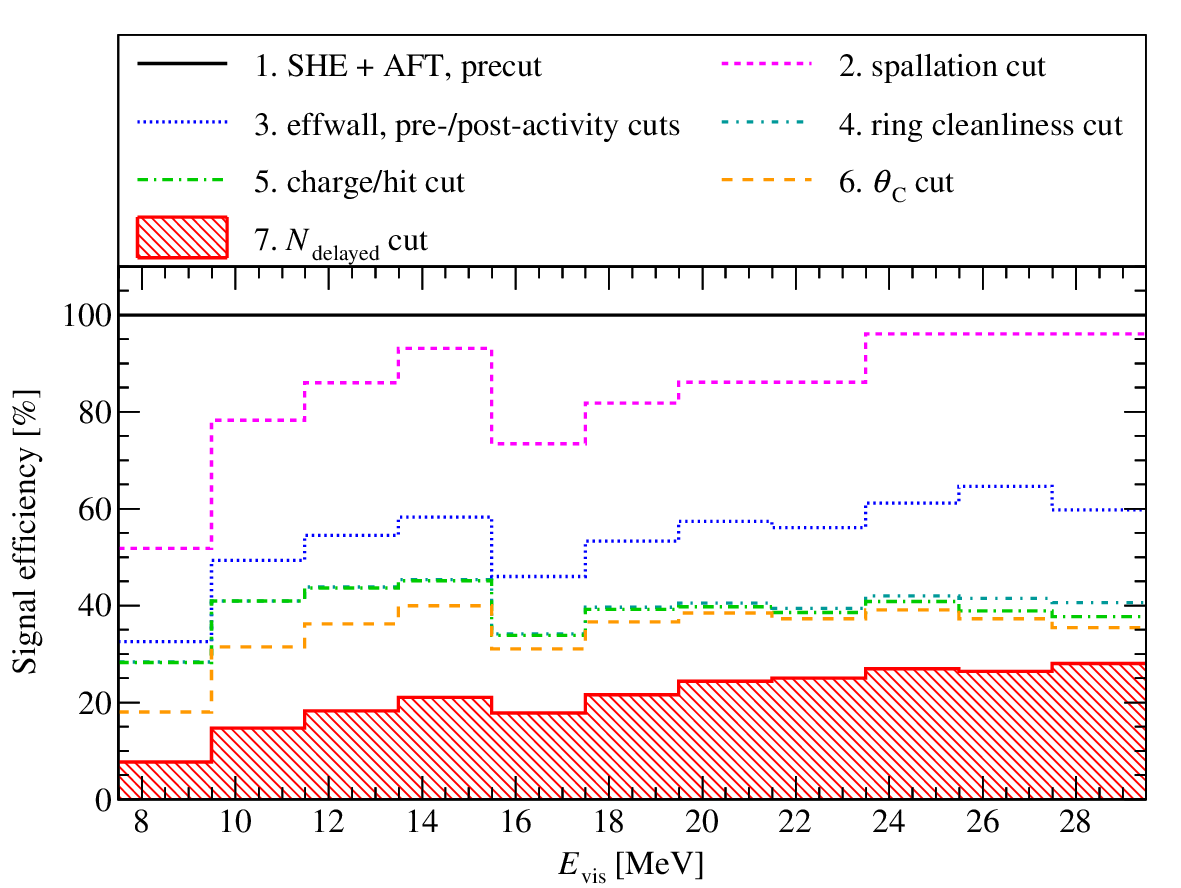}
\caption{\label{SigEff} NCQE cumulative signal efficiencies.
Event reductions are performed in the order shown in the legend, and before the reduction for spallation events is taken as 100\%.
Details of these event reductions are described in Ref.~\cite{2023Harada,2021Abe}.}
\end{figure}

The difference between this study and the DSNB search in SK-VI~\cite{2023Harada} is the cut criteria for $\theta_{\text{C}}$ and $N_{\text{delayed}}$.
In IBD events, only one relativistic positron and one neutron are emitted.
Therefore, $\theta_{\text{C}}$ and $N_{\text{delayed}}$ tend to be about 42~degrees and one, respectively.
In contrast, in an NCQE event, multiple gamma-rays and multiple neutrons are easily emitted.
When multiple gamma-rays are emitted, $\theta_{\text{C}}$ tends to be larger because of the uniform distribution of the hit PMTs.
Therefore, in this study, we select events for which $\theta_{\text{C}}$ is greater than 50~degrees and $N_{\text{delayed}}$ is greater or equal to one.
The cut criteria of $\theta_{\text{C}}$, $E_{\text{vis}}$ and $N_{\text{delayed}}$ are the same as the study in SK pure-water phase~\cite{2019Linyan}.
%The NCQE signal efficiency for $\theta_{\text{C}}$ cut and $N_{\text{delayed}}$ cut is 79.3\% and 51.3\%, respectively.
%While, the IBD signal efficiency for $\theta_{\text{C}}$ cut and $N_{\text{delayed}}$ cut is 6.7\% and 34.9\%, respectively.}
After applying all event selections to 552.2~days of SK-VI data, 38~events remain.
We confirmed that these events are uniformly distributed.
The NCQE cumulative signal efficiencies are shown in Fig.~\ref{SigEff}.\par

This final sample includes not only NCQE events but also other events such as atmospheric neutrino NC non-QE events, atmospheric neutrino charged-current (CC) events, remaining spallation events, reactor neutrino events, and accidental coincidence events.
To determine the final number of NCQE events, it is necessary to estimate the number of all those events.
The expected numbers of atmospheric neutrino events in each secondary interaction model estimated by the simulation are summarized in Table~\ref{tab:num}.
After applying all event selections, NCQE and NC non-QE events account for about 60\% and about 30\% of total events, respectively.
The expected number of spallation, reactor neutrino, and accidental coincidence events, which are calculated by the same method as Ref.~\cite{2023Harada}, are 0.9, 0.1, and 1.6, respectively. 
Note that the number of DSNB events predicted by the Horiuchi~$+$~09 model~\cite{2009Horiuchi}, which is not included in the expected number of events, is 0.1.

\begin{table}[tbp]
\caption{\label{tab:num}
The expected number of atmosphric neutrino events in each secondary interaction model.
The fractions are summarized in parentheses.
The expected number of spallation, reactor neutrino, and accidental coincidence events are common to each model and are described in the text.}
\begin{ruledtabular}
\begin{tabular}{lrrr}
Components&BERT&BIC &INCL++\\
\hline
Total     &46.0&33.7&34.1 \\
\hline
NCQE      &28.7 (62.4\%)&19.8 (58.8\%)&20.2 (59.2\%) \\
NC non-QE &13.3 (28.9\%)&10.2 (30.3\%)&10.1 (29.6\%) \\
CC        &1.4 (3.0\%)  &1.1 (3.3\%)  &1.2 (3.5\%)   \\
\end{tabular}
\end{ruledtabular}
\end{table}

\section{SECONDARY INTERACTION MODELS}\label{SECOND}
\begin{figure*}[tbp]
\includegraphics[width=17.2cm]{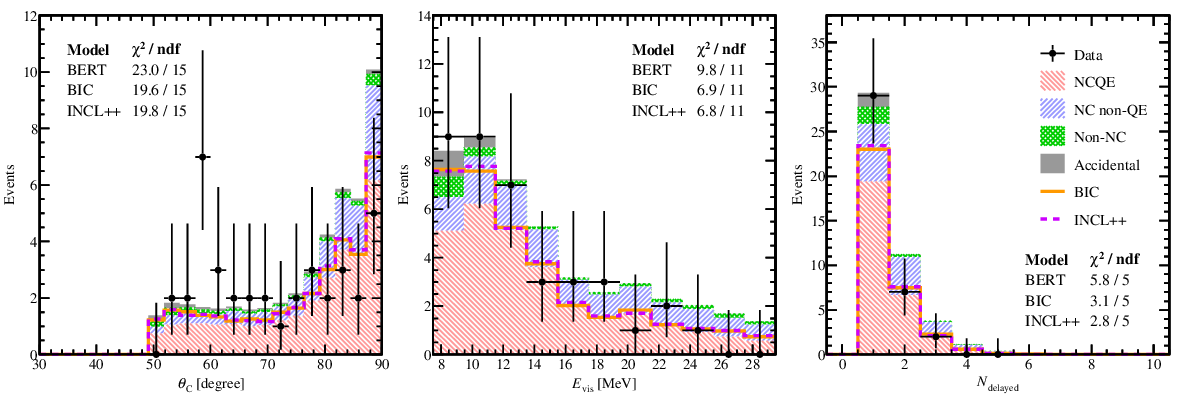}
\caption{\label{Comparison} The distributions of $\theta_{\text{C}}$ (left), $E_{\text{vis}}$ (center), and $N_{\text{delayed}}$ (right).
Each color-filled histogram shows the expected events in BERT.
Non-NC includes CC, spallation, and reactor neutrino events.
Solid and dashed line show the total expected events in BIC and INCL++, respectively.
In all distributions, $\theta_{\text{C}}$ is greater than 50~degrees, $E_{\text{vis}}$ is between 7.49~MeV and 29.49~MeV, and $N_{\text{delayed}}$ is greater or equal to one.
The values of the chi-square are also shown for each distribution.}
%\caption{\label{Comparison} The distributions of $\theta_{\text{C}}$ (left), $E_{\text{vis}}$ (center), and $N_{\text{delayed}}$ (right). Each color-filled histogram shows the expected events in BERT. Non-NC includes CC, spallation, and reactor neutrino events. Solid and dashed line show the total expected events in BIC and INCL++, respectively. In all distributions, $E_{\text{vis}}$ is between 7.49 MeV and 29.49 MeV, and $N_{\text{delayed}}$ is greater or equal to one. The $\theta_{\text{C}}$ (greater than 50 degrees) cut is not applied in the $\theta_{\text{C}}$ distribution only. The values of the chi-square are also shown for each distribution.}
\end{figure*}

In Ref.~\cite{2014Abe,2019Abe}, it was found that agreements of the secondary interaction model based on BERT remain poor and result in significant systematic uncertainty as described in Sec.~\ref{INTROD}.
Therefore, we compare the observed data with the other secondary interaction models using the newly developed Geant4-based SK detector simulation.
Here, we use three secondary interaction models: BERT, BIC, and INCL++.
Fig.~\ref{Comparison} shows the distributions of $\theta_{\text{C}}$, $E_{\text{vis}}$, and $N_{\text{delayed}}$ in each secondary interaction model.\par

The distributions of $\theta_{\text{C}}$, $E_{\text{vis}}$, and $N_{\text{delayed}}$ strongly depend on the number of de-excitation gamma-rays and neutrons.
For example, the direction of Cherenkov photons becomes more isotropic as the number of de-excitation gamma-rays is increased.
Moreover, the total energy of de-excitation gamma-rays is correlated to the number of de-excitation gamma-rays.
Therefore, $\theta_{\text{C}}$ and $E_{\text{vis}}$ become larger as the number of de-excitation gamma-rays gets larger.
Furthermore, $N_{\text{delayed}}$ is correlated to the number of neutrons.\par
In the distributions of $E_{\text{vis}}$ and $N_{\text{delayed}}$, the number of events is larger for BERT than other two models.
Furthermore, in the $\theta_{\text{C}}$ distribution, the differences between BERT and other two models are large in high-angle regions.
These differences come from the number of de-excitation gamma-rays and neutrons by secondary interactions.
The number of de-excitation gamma-rays and neutrons is similar between BIC and INCL++.
While, in BERT, the number of de-excitation gamma-rays and neutrons is larger than in the other two models.\par
We calculated the chi-square $\chi^{\text{2}}$ for $\theta_{\text{C}}$, $E_{\text{vis}}$, and $N_{\text{delayed}}$ distributions by using the Poisson-likelihood~\cite{2020Ji}.
Here, $\chi^{\text{2}}$ is defined as
\begin{eqnarray}
	\chi^{2}=2\sum_{i=\text{1}}^{\text{bin}} \Biggl(N^{\text{exp},i} - N^{\text{obs},i} + N^{\text{obs},i}\,{\rm ln}{N^{\text{obs},i} \over N^{\text{exp},i}}\Biggr),
\end{eqnarray}
where bin is the number of bins, $N^{\text{obs},i}$ is the observed number of events of $i$-th bin and $N^{\text{exp},i}$ is the expected number of events of $i$-th bin.
The values are summarized in Fig.~\ref{Comparison}.
Due to the small statistics, the chi-square cannot give conclusive results; however, the values are smaller for BIC and INCL++ than for BERT in all distributions.\par
As described in Sec.~\ref{INTROD}, an accelerator neutrino-oxygen NCQE cross section measurement was conducted as part of the T2K experiment~\cite{2019Abe}.
The observed and expected number of events in $\theta_{\text{C}}\in$ [78, 90]~degrees and $E_{\text{vis}}\in$ [7.49, 29.49]~MeV obtained in the T2K data analysis are shown in Table~\ref{tab:obsexp}.
We have also performed an analysis using the same criteria of $\theta_{\text{C}}$ and $E_{\text{vis}}$.
The results are also summarized in Table~\ref{tab:obsexp}.
With these selection criteria, the expected number of events in BERT is larger than the observed number of events.
The similar discrepancy was observed in T2K because the secondary interaction model is based on BERT~\cite{2019Abe,1970Coleman}.
In both cases, the differences of the expected and observed number of events in BERT are larger than that in BIC and INCL++, which shows similar trend as above.

\begin{table}[tbp]
\caption{\label{tab:obsexp}
The observed and expected number of events in $\theta_{\text{C}}\in$ [78, 90]~degrees and $E_{\text{vis}}\in$ [7.49, 29.49]~MeV in this study and T2K~\cite{2019Abe}.
In T2K, $N_{\text{delayed}}$ (greater or equal to one) cut is not applied.}
\begin{ruledtabular}
\begin{tabular}{lrrr}
					 &Model &Expected&Observed \\
\hline
					 &BERT  &26.8    &         \\
This study &BIC   &18.4    &14       \\
					 &INCL++&18.9    &         \\
\hline
T2K        &BERT  &100.8   &61       \\
\end{tabular}
\end{ruledtabular}
\end{table}

\section{NCQE CROSS SECTION}\label{NCQECS}
\subsection{Measured NCQE cross section}
The flux-averaged theoretical neutrino-oxygen NCQE cross section is
\begin{eqnarray}
\langle \sigma^{\text{theory}}_{\text{NCQE}} \rangle &=& \frac{\int_{\text{160 MeV}}^{\text{10 GeV}}\sum_{i=\nu,\bar{\nu}}\phi_{i}(E)\times\sigma_{i}(E)^{\text{theory}}_{\text{NCQE}}dE}{\int_{\text{160 MeV}}^{\text{10 GeV}}\sum_{i=\nu,\bar{\nu}}\phi_{i}(E)dE} \nonumber
\\
																										 &=&\text{1.02}\times\text{10}^{-\text{38}}\,\text{cm}^{\text{2}}/\text{oxygen},\label{Eq_theory}
\end{eqnarray}
where $\phi_{i}(E)$ is the atmospheric neutrino flux~\cite{2011Honda} at neutrino energy $E$ and $\sigma_{i}(E)^{\text{theory}}_{\text{NCQE}}$ is the theoretical NCQE cross section~\cite{2012Ankowski}.
The integral is performed between 160~MeV and 10~GeV because the NCQE cross section is small below 160~MeV and the atmospheric neutrino flux is small above 10~GeV.
The systematic uncertainty related to the energy cutoff is described in Sec.~\ref{SYSTEM}.
The measured neutrino-oxygen NCQE cross section is
\begin{eqnarray}
\langle \sigma^{\text{measured}}_{\text{NCQE}} \rangle &=&\frac{N^{\text{obs}}-N^{\text{exp}}_{\text{Non-NCQE}}}{N^{\text{exp}}_{\text{NCQE}}}\times\langle \sigma^{\text{theory}}_{\text{NCQE}} \rangle \nonumber
\\
&=&\text{0.74}\pm\text{0.22}(\text{stat.}) \nonumber
\\
&&\times\,\text{10}^{-\text{38}}\,\text{cm}^{\text{2}}/\text{oxygen},
\end{eqnarray}
where $N^{\text{obs}}$~($=$\,38) is the observed number of events, $N^{\text{exp}}_{\text{NCQE}}$ is the expected number of NCQE events, and $N^{\text{exp}}_{\text{Non-NCQE}}$ is the expected number of non-NCQE events, including NC non-QE, CC, spallation, reactor neutrino, and accidental coincidence.

\subsection{Systematic uncertainties}\label{SYSTEM}
Systematic uncertainties on the expected NCQE, NC non-QE, and CC events are summarized in Table~\ref{tab:sys}.
We follow the estimation methods of measurements in SK pure-water phase and T2K~\cite{2019Linyan,2014Abe,2019Abe}.
The estimation of each systematic uncertainty is described below.\par

\begin{table}[tbp]
\caption{\label{tab:sys}
Systematic uncertainties of the expected NCQE, NC non-QE, and CC events.
NC non-QE and CC cross section are from Ref.~\cite{2014Abe}.
As for the secondary interaction and energy cutoff, only negative values are considered.
Atmospheric neutrino flux, atmospheric neutrino/antineutrino ratio, data reduction, and neutron-tagging uncertainties are common to NC, NC non-QE, and CC and are described in the text.}

\begin{ruledtabular}
\begin{tabular}{lrrr}
                     &  NCQE                                 &  NC non-QE                            &  CC \\
\hline
Cross section        &  -                                    &  $\pm$18.0\%                          &  $\pm$24.0\% \\

%Primary interaction  &  $^{+\text{1.9\%}}_{-\text{9.8\%}}$   &  $^{+\text{0.0\%}}_{-\text{2.6\%}}$   &  $^{+\text{7.3\%}}_{-\text{10.4\%}}$ \\
\multirow{2}{*}{Primary interaction}& $+$1.5\% & $+$0.0\% & $+$1.2\%  \\
                                    & $-$9.4\% & $-$2.4\% & $-$8.0\% \\
Secondary interaction&  $-$30.9\%                            &  $-$24.3\%                            &  $-$20.7\% \\
%\multirow{2}{*}{Secondary interaction}& $+$0.0\%  & $+$0.0\%  & $+$0.0\%  \\
%                                      & $-$32.1\% & $-$22.9\% & $-$20.9\% \\
Energy cutoff        &  $-$2.1\%                             &  $-$1.5\%                             &  $-$19.9\% \\
\end{tabular}
\end{ruledtabular}
\end{table}

Primary interaction uncertainty arises from the spectroscopic strengths of the oxygen nucleus.
Computation of the $p_{\text{3}/\text{2}}$ spectroscopic strength is consistent with $^{\text{16}}\text{O}(\text{e}, \text{e}'\text{p})$ experiment within 5.4\%~\cite{2012Ankowski,1994Leuschner}.
For the \textit{others} state, there is no reliable predictions as written in Sec.~\ref{SIMULA}, thus the uncertainty is conservatively estimated by comparing with an extreme case, that is the difference between the default state ($(s_{\text{1}/\text{2}})^{-\text{1}}$) and the ground state ($(p_{\text{1}/\text{2}})^{-\text{1}}$).\par
Secondary interaction uncertainty arises from the secondary interaction model used in the detector simulation.
As described in Sec.~\ref{SECOND}, the chi-square differences were inconclusive.
Therefore, the uncertainty is taken to be the difference in the expected number of events from BERT to BIC or INCL++.\par
In the calculation of Eq.~(\ref{Eq_theory}), the integral is performed between 160~MeV and 10~GeV, while the expected number of atmospheric neutrino events is estimated using full energy range.
Energy cutoff uncertainty is estimated by applying the energy cutoff to the expected number of atmospheric neutrino events.
Since the expected number of events decreases by the energy cutoff, only negative systematic uncertainty is considered.\par
The uncertainty of the measured atmospheric neutrino flux in SK differs in each energy bin, as shown in Fig.~6 and Table~IV of Ref.~\cite{2016Richard}.
In this measurement, we chose the conservative value and 18.0\% in [160~MeV, 10~GeV] is applied to atmospheric neutrino flux uncertainty.\par
Atmospheric neutrino/antineutrino ratio, data reduction, and neutron-tagging uncertainties are taken as 5.0\%~\cite{2007Honda}, 1.4\%~\cite{2021Abe}, and 6.4\%~\cite{2023Harada}, respectively.\par
Systematic uncertainties of spallation, reactor neutrino, and accidental coincidence events are taken as 60.0\%~\cite{2023Harada}, 100.0\%~\cite{2023Harada}, and 4.6\%, respectively.
Due to the small event fraction, these uncertainties are negligible.\par

\begin{figure}[tbp]
\includegraphics[width=8.6cm]{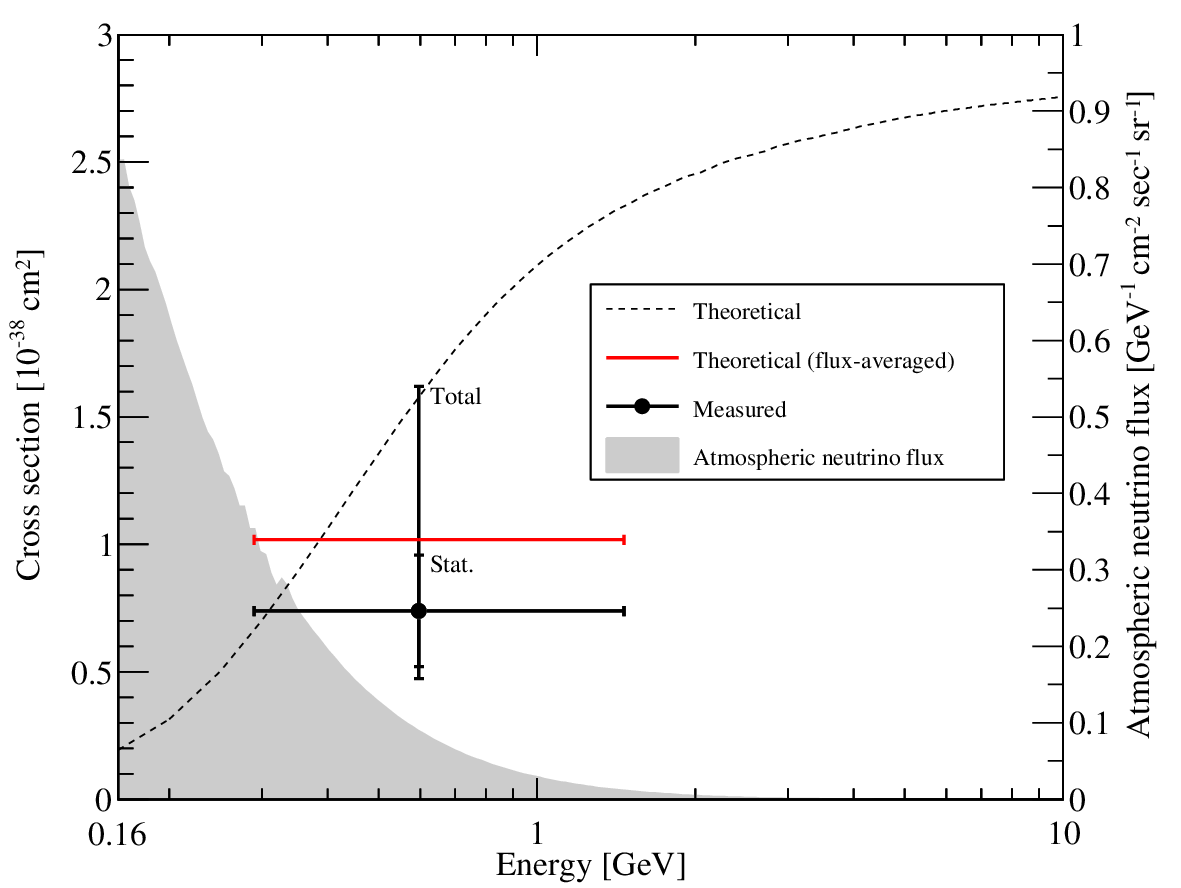}
\caption{\label{MeasuredNCQE} The measured neutrino-oxygen NCQE cross section, the theoretical neutrino-oxygen NCQE cross section~\cite{2012Ankowski}, and the atmospheric neutrino flux predicted using the HKKM11 model~\cite{2011Honda}.
%The atmospheric neutrino flux is shown with an arbitrary normalization.
%Vertical bars show the statistical uncertainty (short bar) and the quadratic sum of statistical and systematic uncertainties (long bar).
Vertical bars show the statistical uncertainty (short bar) and the total uncertainty (long bar).
Horizontal bars show the 1$\sigma$ from the mean (0.60~GeV) of the theoretical NCQE cross section multiplied by the atmospheric neutrino flux.}
\end{figure}

Systematic uncertainty of the measured NCQE cross section is estimated by performing toy MC considering the systematic uncertainties.
As a result, the 1$\sigma$ confidence level region becomes [0.59, 1.59]\,$\times\,\text{10}^{-\text{38}}\,\text{cm}^{\text{2}}$$/$oxygen, and the measured NCQE cross section is determined as
\begin{eqnarray}
\langle \sigma^{\text{measured}}_{\text{NCQE}} \rangle &=&\text{0.74}\pm\text{0.22}(\text{stat.})^{+\text{0.85}}_{-\text{0.15}}(\text{syst.}) \nonumber
\\ 
&&\times\,\text{10}^{-\text{38}}\,\text{cm}^{\text{2}}/\text{oxygen}.
\end{eqnarray}
The measured NCQE cross section, the theoretical NCQE cross section~\cite{2012Ankowski}, and the atmospheric neutrino flux predicted using the HKKM11 model~\cite{2011Honda} are shown in Fig.~\ref{MeasuredNCQE}.
The measured NCQE cross section is consistent with the flux-averaged theoretical NCQE cross section within the uncertainties.
Furthermore, the measured NCQE cross section is consistent with the measurement in the SK pure-water phase within the uncertainties (1.01\,$\pm$\,0.17(stat.)\,$^{+\text{0.78}}_{-\text{0.30}}$(syst.)\,$\times$\,10$^{-\text{38}}$\,cm$^{\text{2}}$$/$oxygen)~\cite{2019Linyan}.
The systematic uncertainty on the measured NCQE cross section in this study is larger than that in the measurement of the SK pure-water phase.
The reason is that we take the difference of secondary interaction models into consideration, conservatively estimated by the comparison among these models.
The uncertainty will be reduced with better understanding of secondary interaction models in future.

\section{CONCLUSION AND FUTURE PROSPECTS}\label{CONCLU}
We reported the first measurement of the atmospheric neutrino-oxygen NCQE cross section in the Gd-loaded SK water Cherenkov detector.
Using a 552.2~day data set, the NCQE cross section was measured to be 0.74\,$\pm$\,0.22(stat.)\,$^{+\text{0.85}}_{-\text{0.15}}$(syst.)\,$\times$\,10$^{-\text{38}}$\,cm$^{\text{2}}$$/$oxygen in the energy range from 160~MeV to 10~GeV, which was consistent with the atmospheric neutrino-flux-averaged theoretical NCQE cross section (1.02\,$\times$\,10$^{-\text{38}}\,$cm$^{\text{2}}$$/$oxygen) and the measured NCQE cross section in the SK pure-water phase (1.01\,$\pm$\,0.17(stat.)\,$^{+\text{0.78}}_{-\text{0.30}}$(syst.)\,$\times$\,10$^{-\text{38}}$\,cm$^{\text{2}}$$/$oxygen).
Moreover, from the comparison of three different secondary interaction models, we found that BIC and INCL++ provide a somewhat better fit to the observed data than BERT.\par
As described in Sec.~\ref{SK}, we continue the observation with a 0.03\% Gd-loaded SK detector, the phase known as SK-VII.
Since the neutron-tagging efficiency in SK-VII is higher than that in SK-VI~(35.6\%)~\cite{2023Harada,2022Harada}, more delayed signals can be detected, and the observed number of events can be accumulated faster in SK-VII than in SK-VI.
Assuming that the neutron-tagging efficiency in SK-VII is about 60\%, the statistics increases by about 1.4 times with the same live time as SK-VI.
After one more year of observation in SK-VII the statistical uncertainty will reach the NCQE cross section measurement in the SK pure-water phase, and the secondary interaction models will be able to be verified more precisely.
Additional measurement using T2K's accelerator neutrino beam interactions in SK-Gd will help to further refine the physics models for the secondary interactions.

\begin{acknowledgments}
% put your acknowledgments here.
We gratefully acknowledge the cooperation of the Kamioka Mining and Smelting Company.
The Super-Kamiokande experiment has been built and operated from funding by the Japanese Ministry of Education, Culture, Sports, Science and Technology, the U.S. Department of Energy, and the U.S. National Science Foundation.
Some of us have been supported by funds from the National Research Foundation of Korea (NRF-2009-0083526 and NRF 2022R1A5A1030700) funded by the Ministry of Science, Information and Communication Technology (ICT), the Institute for Basic Science (IBS-R016-Y2), and the Ministry of Education (2018R1D1A1B07049158, 2021R1I1A1A01042256),
the Japan Society for the Promotion of Science, including KAKENHI (JP22KJ2301),
the National Natural Science Foundation of China under Grants No. 11620101004,
the Spanish Ministry of Science, Universities and Innovation (grant PGC2018-099388-B-I00),
the Natural Sciences and Engineering Research Council (NSERC) of Canada, the Scinet and Westgrid consortia of Compute Canada,
the National Science Centre (UMO-2018/30/E/ST2/00441) and the Ministry of Education and Science (DIR/WK/2017/05), Poland,
the Science and Technology Facilities Council (STFC) and Grid for Particle Physics (GridPP), UK,
the European Union's Horizon 2020 Research and Innovation Programme under the Marie Sklodowska-Curie grant agreement no. 754496,
%the European Union's Horizon 2020 Research and Innovation Programme under the Marie Sk\l odowska-Curie grant agreement no. 754496,
H2020-MSCA-RISE-2018 JENNIFER2 grant agreement no. 822070, and 
H2020-MSCA-RISE-2019 SK2HK grant agreement no. 872549.
\end{acknowledgments}

% Create the reference section using BibTeX:
%\nocite{*}
\bibliography{Sakai.bbl}
%\bibliography{SakaiBiB}

\end{document}